\documentclass[conference]{IEEEtran}
\IEEEoverridecommandlockouts
\ifCLASSINFOpdf
\else
\fi

\usepackage{amssymb}
\usepackage{amsmath}
\usepackage{amsthm}
\usepackage{graphicx}
\usepackage{algorithm}
\usepackage{algorithmic}
\usepackage{booktabs}
\usepackage{cite}
\usepackage{bm}
\usepackage{float}
\usepackage{multirow}
\usepackage{color}
\usepackage{subfigure}
\usepackage{caption}
\usepackage{epstopdf}
\usepackage{hyperref}
\usepackage{fixltx2e}
\usepackage{longtable}
\usepackage{diagbox}
\usepackage{changepage}
\usepackage{comment}
\usepackage{cases}
\usepackage[short]{newoptidef}
\usepackage{makecell}
\usepackage{mathabx}
\theoremstyle{definition}

\theoremstyle{remark}

\theoremstyle{plain}

\newcommand{\tmop}[1]{\ensuremath{\operatorname{#1}}}

\begin{document}
		%
		\title{Learning-based Sustainable Multi-User Computation Offloading for Mobile Edge-Quantum Computing
		\author{Minrui~Xu, Dusit~Niyato, \emph{Fellow,~IEEE}, Jiawen~Kang, \emph{Member,~IEEE}, Zehui Xiong, \emph{Member,~IEEE},\\ and Mingzhe~Chen,~\emph{Member, IEEE}}
    \thanks{M.~Xu and D.~Niyato are with the School of Computer Science and Engineering, Nanyang Technological University, Singapore (e-mail: minrui001@e.ntu.edu.sg; dniyato@ntu.edu.sg). J.~Kang is with the School of Automation, Guangdong University of Technology, China (e-mail: kavinkang@gdut.edu.cn). Z.~Xiong is with the Pillar of Information Systems Technology and Design, Singapore University of Technology and Design, Singapore 487372, Singapore (e-mail: zehui\_xiong@sutd.edu.sg). M.~Chen is with the Department of Electrical and Computer Engineering and Institute for Data Science and Computing, University of Miami, Coral Gables, FL, 33146 USA (e-mail: mingzhe.chen@miami.edu). 
    }
 }
		\maketitle
		\begin{abstract}
			
  In this paper,
  a novel paradigm of mobile edge-quantum computing (MEQC) is proposed, which brings quantum computing capacities to mobile edge networks that are closer to mobile users (i.e., edge devices). First, we propose an MEQC system model where mobile users can offload computational tasks to scalable quantum computers via edge servers with cryogenic components and fault-tolerant schemes. Second, we show that it is NP-hard to obtain a centralized solution to the partial offloading problem in MEQC in terms of the optimal latency and energy cost of classical and quantum computing.
		Third, we propose a multi-agent hybrid discrete-continuous deep reinforcement learning using proximal policy optimization to learn the long-term sustainable offloading strategy without prior knowledge. Finally, experimental results demonstrate that the proposed algorithm can reduce at least 30\% of the cost compared with the existing baseline solutions under different system settings.
		\end{abstract}
		

		%
		\IEEEpeerreviewmaketitle

		\section{Introduction}
		
		With its significant economic, environmental, and societal opportunities, quantum computing is regarded as a strategic technology for academic and industrial domains. It has a tremendous bidirectional impact on existing technology fields such as artificial intelligence (AI)~\cite{biamonte2017quantum}, security~\cite{bharti2021noisy}, and finance~\cite{herman2022survey}. For example, quantum computers can be used for molecular simulations and discoveries in materials science and biology to develop new materials and drugs. In addition, quantum computers can break the encryption, e.g., RSA and secret key exchanging~\cite{herman2022survey}, used to secure modern digital communications, blockchains, and cryptocurrencies. However, performing these quantum computational tasks requires scalable quantum computers with around 10$^6$ physical qubits.
		
		Different from classical mobile edge computing (MEC) using classical information technologies~\cite{chen2015efficient}, as the number of qubits, gates, and measurements in scalable quantum computers increases, the inherent qubit noise will affect  qubit fidelity~\cite{resch2021benchmarking}. To achieve fault-tolerant quantum computing, scalable quantum computers can operate quantum processors with advanced cryogenic components and fault-tolerant schemes~\cite{martin2022energy}. On the one hand, quantum computers work at extremely low temperatures to cool quantum devices to a low-entropy state. On the other hand, quantum noise is combated through the use of fault-tolerant schemes, including those which employ concatenated codes, surface codes, and bosonic qubits~\cite{resch2021benchmarking}. For instance, in error-correction codes, the information is distributed across multiple physical qubits that constitute a logical qubit. Overall, tremendous improvements in computing and energy advantages of scalable quantum computers can be achieved only with fault-tolerant quantum computing operating at cryogenic temperatures.
		

		\begin{figure}
		    \centering
		    \includegraphics{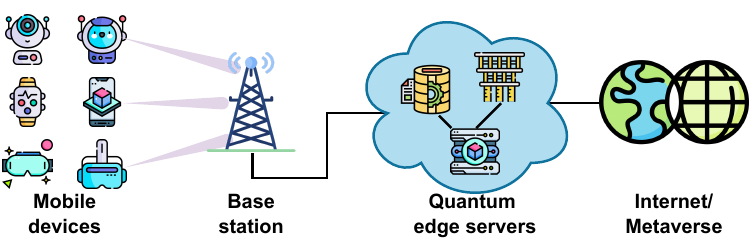}
		    \caption{An illustration of mobile edge-quantum computing.}
		    \label{fig:system}
		\end{figure}
		When scalable quantum computing reaches the scale and quality required, mobile edge-quantum computing (MEQC), combined with the remote access that many edge servers provide, will continue to push the boundaries of quantum advantage. By offloading computational tasks to quantum computers in edge servers, users around the world can obtain the possibility to experience and benefit from cloud/edge quantum computer providers, such as Amazon Braket\footnote{\url{https://aws.amazon.com/braket}}, IBM Quantum\footnote{\url{https://www.ibm.com/quantum}}, and Azure Quantum\footnote{\url{https://azure.microsoft.com/en-us/products/quantum}}. This extends the potential of discovering innovative applications for quantum computing and tackling existing issues, thus driving the practical implementation of quantum computing.
		MEQC can provide users with a range of potentially lethal applications, such as quantum ray tracing~\cite{santos2022towards}, by significantly increasing the appeal of quantum computing to mobile consumers. Unlike traditional edge cloud computing~\cite{chen2015efficient}, MEQC has significant differences in terms of computing power and energy consumption. First, quantum computing uses the superposition, interference, and entanglement of quanta to accelerate computational tasks differently. 
        Second, quantum computers generally operate in an extremely low-temperature environment. Compared to conventional computers, most of the energy consumption of quantum computers is used to maintain the ultra-low temperature container. Third, for the reliability of computational results, quantum computers choose appropriate error correction codes and the concatenation degree of error correction according to their energy and latency constraints. Therefore, enabling mobile users (i.e., edge devices) to offload tasks efficiently to quantum edge servers is still a challenging problem.
		
		The problem of finding sustainable offloading computation strategies in MEQC involves continuous and discrete optimization variables in both classical and quantum computing. Therefore, deciding on how much computation and which server to offload is a complex mixed-integer programming problem that is NP-hard. Fortunately, deep reinforcement learning (DRL)~\cite{neunert2020continuous, schulman2017proximal} is used by mobile users to learn an optimal offloading strategy without prior knowledge of the computational tasks and local environment. With the goal of maximizing long-term returns, DRL can be used to learn the optimal offloading strategy in the presence of dynamically changing quantum noise and the cost of computational tasks. In this way, mobile users can learn to determine the allocation profile and offloading ratio for edge quantum computing to maximize the utilization of MEQC.
		
		The main contributions can be summarized as follows.
		\begin{itemize}
		    \item We propose a novel paradigm of mobile edge-quantum computing that brings quantum advantages from the cloud to mobile users. We propose a multi-user multi-server MEQC system where users can dynamically and flexibly offload partial  computational tasks to servers.
		    \item Based on the MEQC system model, we formulate a classical and quantum computation offloading problem as a mixed integer programming model.
		    \item To improve the offloading efficiency in MEQC, we formulate the multi-user multi-server partial computation offloading problem as a partially observable Markov decision process (POMDP) and propose a multi-agent hybrid continuous-discrete DRL algorithm to learn the optimal sustainable offloading strategy without prior knowledge.
		\end{itemize}
		
		
\section{System Model and Problem Formulation}
  We consider an MEQC system that consists of a set $\mathcal{U}$ of $U$ mobile users and a set $\mathcal{E}$ of $E$ quantum edge servers. We assume that each user $u$ requests a computational task $\mathcal{T}_u^\emph{C} \triangleq (s_u,n_u)$ that can be executed partially by local central processing units (CPUs) and one edge server that has both CPUs and quantum processing units (QPUs). Here, $s_u$ represents the data size of each raw computational task
  $\mathcal{T}_u^\emph{C}$ and $n_u$ represents the required CPU cycles per data size to accomplish computational task $\mathcal{T}_u^\emph{C}$. Next, we first introduce the delay and energy that each device processes a partial computational task with its local CPUs.

  \subsection{Local Computing Model}
  Here, we first introduce the latency of each user processing a partial computational task.
  First, we assume that the computational capacity of each user $u$ is $f_u^L$ (i.e., CPU cycles per second) and the proportion of a task that user $u$ processes locally is $\phi_u$. 
  Then, the latency of user $u$ processing a partial task $\phi_u s_u$ locally is $d_u^L(\phi_u) = \frac{\phi_u s_u n_u}{f_u^L}$. Meanwhile, the energy consumption of user $u$ processing a partial task $\phi_u s_u$ is $e^L_u(\phi_u) = \gamma \phi_u s_u n_u$, where $\gamma$ is the chip coefficient.
  The cost of user $u$ processing partial computational task $\phi_u s_u$ is
		\begin{equation}
		    c_u^L(\phi_u) = \lambda_u^\emph{D} d_u^L(\phi_u)+ \lambda_u^\emph{E} e^L_u(\phi_u),
		\end{equation}
		where $\lambda_u^\emph{D}, \lambda_u^\emph{E}\in [0, 1]$ denote the weight parameters of serving latency and energy, respectively. 
  
  \subsection{Edge Computing Model}
  Next, we introduce the energy and delay that each user offloads its computational task $\left(1-\phi_u\right)s_u$  to one edge server. Here, task offloading consists of the computational task transmission phase and the computational processing phase. During the task processing phase,  each server can use either CPUs or quantum computing to process its received computational task. We assume that each server can use its quantum computing to process only one task per time~\cite{resch2021benchmarking}.

\subsubsection{Task Transmission}
  The time and energy that user $u$ uses to transmit the data with size $\left(1-\phi_u\right)s_u$ are given by
    \begin{equation}\label{eq:transmissiond}
        d_u^\emph{O}(\phi_u,\boldsymbol{a}) = \frac{(1-\phi_u)s_u}{r_u(\boldsymbol{a})},
    \end{equation}
    and
    \begin{equation}
        e_u^\emph{O}(\phi_u,\boldsymbol{a}) = \frac{p_u (1-\phi_u) s_u}{r_u(\boldsymbol{a})},
    \end{equation}
  respectively. Here, $r_u(\boldsymbol{a}) = B \log_2 (1 + \frac{p_u g_{u}(a_u)}{\sigma_{a_u}^2})$ is the uplink data rate from user $u$ to server $a_u$ with $B$ being the bandwidth, $g_{u}(a_u)$ being the channel gain between user $u$ and edge server $a_u$, $\sigma_{a_u}$ being the AGWN noise at server $a_u$, and $p_u$ being the transmit power of user $u$.

\subsubsection{Task Processing at Edge Server} When a server receives the computational task from $u$, it needs to determine whether to use CPUs or quantum computing to process the task. Next, we first introduce the energy and time that the server uses CPUs to process the task of each user $u$. 
\paragraph{CPUs for Task Processing} Let $f_u^E$ be the computing resource that a server uses to process the task of user $u$. 
We assume that the subscribed computing resources should always be satisfied since the Internet/Metaverse operator can invest in the large-scale edge computing infrastructure~\cite{chen2015efficient}
  Then, the time of server $a_u$ processing user $u$'s partial offloaded task with size $\left(1-\phi_u\right)s_u$ is
		\begin{equation}
		    d^\emph{E}_u(\phi_u) = \frac{(1-\phi_u) s_u n_u}{f^E_u}.
		\end{equation}
The energy consumption that server $a_u$ uses CPUs to process user $u$'s partial offloaded task with size $\left(1-\phi_u\right)s_u$ is 
		\begin{equation}\label{eq:energycomputation}
		    e^\emph{E}_u(\phi_u) = \gamma (1-\phi_u) s_u n_u.
		\end{equation}
Given (\ref{eq:transmissiond})-(\ref{eq:energycomputation}), the total cost of a server processing user $u$'s offloaded task with size $\left(1-\phi_u\right)s_u$ is expressed by 
\begin{equation}
\begin{aligned}
    c_u^E(\phi_u,\boldsymbol{a}) = &\lambda_u^\emph{D} \left[d_u^\emph{O}(\phi_u,\boldsymbol{a})+d_u^\emph{E}(\phi_u)\right] \\ & + \lambda_u^\emph{E} \left[e_u^\emph{O}(\phi_u,\boldsymbol{a})+e_u^\emph{E}(\phi_u)\right].
\end{aligned}
\end{equation}



\paragraph{QPUs for Task Processing}
  Here, we first introduce the process of using quantum computing to process a computational task. Then, we model the energy and latency that each server uses quantum computational resources for task processing. 
One quantum computer containing many physical qubits located at each edge server operates in an extremely low-temperature environment. Each edge server can transform one of its received computational task into a quantum circuit by quantum algorithms to perform quantum acceleration for the task. In particular, let $\mathcal{T}^\emph{Q}_u = (s_u, Q_u, D_u)$ be the quantum computational task transformed from the computational task of user $u$ where $Q_u$ is the required number of qubits to execute the quantum circuit, and $D_u$ is the length of the quantum circuit. An exemplary application of using quantum computing to process a computational task is discussed in~\ref{sec:application}.

To construct scalable quantum computers, error correction schemes leverage multiple noisy qubits to constitute a single logical qubit with high fidelity.
Consequently, error correction schemes require many gate operations, and hence its energy power consumption is nearly independent of the used quantum algorithm actually having at the logical level~\cite{fellous2022optimizing}. 
Let $N_{1}$, $N_{2}$, and $N_\textrm{M}$ be the average numbers of physical one-qubit (1qb) gates, two-qubit (2qb) gates, and measurement gates, respectively, run in parallel per time step of the circuit. The cost of running quantum computers in MEQC consists of latency and energy costs. Specifically, the latency in quantum computing is mainly caused by the operation of logical gates in quantum circuits~\cite{fellous2022optimizing}, which can be calculated as
\begin{equation}
    d_u^\emph{Q}(\phi_u) = (1-\phi_u)s_u Q_u \big[\tau_{1}N_{1} + \tau_{2}N_{2} + \tau_\textrm{M}N_\textrm{M}\big],
\end{equation}
where $\tau_{1}$, $\tau_{2}$, and $\tau_\textrm{M}$ are the time of processing 1qb gates, 2qb gates, and measurement gates, respectively.

Quantum computers are operated at a cryogenic temperature for the low entropy state, and thus initial states of qubits can be prepared accurately. Therefore, the energy of quantum computing is mainly caused by the cooling system used to maintain the low temperature~\cite{martin2022energy}. The energy consumption of quantum computers~\cite{fellous2022optimizing} in edge servers can be defined as
\begin{equation}
    e_u^\emph{Q}(\phi_u) = (1-\phi_u) s_u Q_u \big[P_{1}N_{1} + P_{2}N_{2} + P_\textrm{M}N_\textrm{M} + P_\textrm{Q} Q\big],
\end{equation}
where $Q$ is the number of physical qubits in one logical qubit, $P_1, P_2, P_\textrm{M}$, and $P_\textrm{Q}$ are energy consumption of each physical 1qb gate, 2qb gate, measurement gate, and qubit, respectively. The specific equations of $P_1, P_2, P_\textrm{M}$, and $P_\textrm{Q}$ are shown in Appendix~\ref{app:energy}. 

The total cost of a server using quantum computing to process user $u$'s offloaded task in terms of running time and energy of quantum processors is
\begin{equation}
\begin{aligned}
    c_u^\emph{Q}(\phi_u,\boldsymbol{a}) = &\lambda_u^\emph{D} \left[d_u^\emph{O}(\phi_u,\boldsymbol{a})+d_u^\emph{Q}(\phi_u)\right] \\&+ \lambda_u^\emph{E} \left[e_u^\emph{O}(\phi_u,\boldsymbol{a})+e_u^\emph{Q}(\phi_u)\right].
\end{aligned}
\end{equation}

Although ultra-low temperature environments and error correction schemes are used to improve the scalability of quantum computing, the inherent noise in quantum circuits cannot be completely eliminated~\cite{krinner2022realizing}. Therefore, the success probability is used to describe the performance of quantum circuits under nondeterministic quantum operations. As the width and depth of quantum circuits increase, the number of locations where errors can occur in quantum circuits also increases, which leads to a lower success probability of edge quantum computing~\cite{resch2021benchmarking}. By employing the common noise model discussed and error probability $\epsilon_{\textrm{err}}$ per physical gate calculated in Appendix~\ref{app:noise}, the linear approximation of the success probability is~\cite{fellous2022optimizing}
\begin{equation}
  \mathcal{M}_u(a_u) = 1 - \mathcal{N}^L_u \epsilon_{\tmop{thr}} (\epsilon_{\tmop{err}}
  / \epsilon_{\tmop{thr}})^{2^{k_{a_u}}},
\end{equation}
where $\mathcal{N}^L_u = Q^L_u \times D^L_u$ denotes the number of locations where
logical errors can happen, $\epsilon_{\tmop{thr}}$ is the threshold for error
correction, $\epsilon_{\tmop{err}}$ is the errors per physical gate, and $k_{a_u}$ is the error correction's concatenation level at edge server $a_u$.

Let $Q_u^E$ denote quantum computing capacity (i.e., number of logical qubits) at edge servers subscribed by user $u$ from the Internet/Metaverse operator. Therefore, we can have an indicator function $I_u(\boldsymbol{a})$ with $I_u(\boldsymbol{a})=1$ means that the task can be executed by quantum computers at $a_u$ and $I_u(\boldsymbol{a})=0$ otherwise, which can be expressed as 
\begin{equation}
    I_u(\boldsymbol{a}) = \begin{cases}
1, & \text{ if } Q_u \leq Q_u^E \quad \&\quad \mathcal{M}(a_u) \geq 2/3, \\ 
0, & \text{ otherwise.}
\end{cases}
\end{equation}
Here, the threshold success probability of quantum circuits is set to 2/3, which is a classical choice for a single run of a quantum algorithm~\cite{fellous2022optimizing}.
  \subsection{Problem Formulation}
    Overall, based on the above local computing model and edge computing model, the total offloading and execution cost in MEQC can be calculated as
  \begin{equation}
  \begin{aligned}
      C(\boldsymbol{a}, \boldsymbol{\phi}) = \sum_{u\in\mathcal{U}} \bigg[c_u^L(\phi_u)  + (1&-I_u(\boldsymbol{a})) c_u^\emph{E}(\boldsymbol{a}, \phi_u) \\& + I_u(\boldsymbol{a}) c_u^\emph{Q}(\boldsymbol{a}, \phi_u)\bigg].
  \end{aligned}\label{eq:cost}
  \end{equation}
    Given Eq.~\eqref{eq:cost}, the problem of minimizing the total cost of task offloading in MEQC can be formulated as
\begin{mini!}|s|[2]<b>
  {\boldsymbol{a}, \boldsymbol{\phi}}{C(\boldsymbol{a}, \boldsymbol{\phi}),}{\label{obj}}{}
  \addConstraint{\sum_{u\in \mathcal{U}} \boldsymbol{1}_{\{a_u=e\}}I_u(\boldsymbol{a})}{\leq 1,}{\quad e\in \mathcal{E} \label{con0}}{}
  \addConstraint{a_u}{\in \{1,\ldots,E\},}{ \quad \forall u \in \mathcal{U}\label{con1}}{}
  \addConstraint{\phi_{u}}{\in [0, 1],}{\quad \forall u \in \mathcal{U}.\label{con2}}{}
    \end{mini!}
    The constraint in Eq. \eqref{con0} means that each edge server can only perform one quantum computing task from MEC. Moreover, the constraint in Eq. \eqref{con1} means that each user can only select one of the edge servers to offload. Finally, the constraint in Eq. \eqref{con2} represents that the partial offloading ratio is between 0 and 1. 

 \section{A Exemplary Application of Mobile Edge-Quantum Computing}\label{sec:application}
In the aforementioned general system model, a classical computational task $\mathcal{T}_u^\emph{C}$ of user $u$ can be compiled into a quantum computation task $\mathcal{T}_u^\emph{Q}$ and executed by quantum computers at edge servers. Here, we provide an exemplary application of MEQC, i.e., quantum ray tracing, which is generated by users and can be executed via local computing, edge classical computing, or edge quantum computing~\cite{santos2022towards}.
		
\subsection{Basics of Quantum Computing}\label{sec:basics}
By leveraging quantum mechanics, including entanglement, parallelism, and  superposition, information of classical bits can be encoded into quantum bits, or qubits, which can be not only in the state $|0\rangle$ and $|1\rangle$ but also their superposition $\alpha|0\rangle+\beta|1\rangle$, where $\alpha,\beta\in\mathbb{C}$ and $|\alpha|^2 + |\beta|^2 = 1$.
The superposition of $n$ qubits can be represented by $|b\rangle^{\otimes n} = \sum_{i=0}^{2^n-1}\alpha_i|i\rangle$, where $\forall \alpha_i\in\mathbb{C}$ and $\sum_{i=0}^{2^n-1}|\alpha_i|^2 = 1$. Based on the superposition, the system can represent $N = 2^n$ states simultaneously with $n$ qubits. This provides quantum advantages to computation due to exponential quantum parallelism. Similar to classical computing, the manipulation of qubits is achieved by quantum gates, including unitary gates and measurement gates. In detail, unitary gates implement unitary transformations of quantum states and measurement gates implement probabilistic and destructive transformations for classical information extraction from quantum states. Building on qubits and quantum gates, various quantum circuits can be designed to implement corresponding quantum algorithms to achieve effective acceleration of classical computation tasks.

		
		\subsection{Quantum Ray Tracing for the Internet/Metaverse}
		Metaverse~\cite{xu2022full}, as the 3D Internet, allows users to immerse in virtual worlds rendered by computers. Specifically, ray tracing is the primary algorithm behind most rendering algorithms for generating realistic imagery in the Metaverse. However, for the scene with the number of primitives $N$, classical ray tracing algorithms have the complexity of $\mathcal{O}(N)$ for the number of intersection evaluations. By using quantum searching algorithms, such as Grover's algorithm, the complexity of quantum ray tracing algorithms can be reduced to $\mathcal{O}(\sqrt{N})$. 
		
		In the quantum ray tracing algorithm proposed in~\cite{santos2022towards}, the depth range is set for the possible intersections. The depth in rendering algorithms is regarded as the distance from the intersection point to the origin of the ray. With the objective of minimum depth, the ray tracing algorithm attempts to find the primitive that is intersected the closest point to the origin of the ray. Therefore, the depth range is assigned according to the parameter of depth and to the near and far fields of the ray. According to the current ray, the scene, and the depth range, the quantum ray tracing algorithm then compiles the $\hat{R}_r$ operator's quantum circuit. To estimate which primitives among a collection of $P$ primitives, $P=2^{pb}$, $pb\in \mathbb{N}$, are intersected by ray $r$, the ray tracing intersection operator $\hat{R}_r$ uses two quantum registers, i.e., the primitive register $p_{reg}$ and the indication register $i_{reg}$. The primitive register $p_{reg}$ listing all $P$ primitives is initially prepared as a uniform superposition. In addition, the indication register $i_{reg}$ is then changed to $|1\rangle$ if the $r$ is intersected with the primitive $p$ in the superposition:
		\begin{equation}
		    \hat{R}_{r} |0\rangle^{\otimes p b}|0\rangle \mapsto \frac{1}{\sqrt{P}} \sum_{p=0}^{P-1}\big[|p\rangle\left(\left(1-i_{r}(p)\right)|0\rangle+i_{r}(p)|1\rangle\right)\big].
		\end{equation}
		
		Using the Hadamard gates with $pb$ qubits denoted as $\hat{\mathcal{H}}_{pb}$, the superposition on $p_{reg}$ is prepared. The operator $\hat{Int}_r$ implements, for all primitives $p$ indexed by $p_{reg}$, the intersecting function $i_r(p)$ can be defined by
		\begin{equation}
		    i_{r}(p)=\begin{cases}
1, & \text { if } p \text { is intersected the ray } r, \\
0, & \text { otherwise},
\end{cases}
		\end{equation}
		and $\widehat{R}_{r}=\widehat{I n t}_{r}\left(\widehat{\mathcal{H}}_{pb} \otimes \widehat{\mathcal{I}}_{1}\right)$, where $\widehat{\mathcal{I}}_{1}$ represents the identity operator adopted to the qubit in $i_{reg}$.
		Subsequently, to search for a feasible intersection within the possible range of depth, $\hat{R}_r$ as the oracle calls the quantum search algorithm~\cite{herman2022survey}. 
  At termination, the quantum ray tracing algorithm returns whether a valid intersecting primitive was discovered and additional information about that intersection (e.g., primitive ID, normal, 3D point, and depth are all integers).
	Overall, the quantum ray tracing algorithm requires $qb + 2\times cb + 5$ qubits to run and the depth of the circuit is $3 \times qb + \mathcal{O}( \lfloor  \frac{\pi}{4} \sqrt{2^{qb + 2\times cb + 5}} \rfloor)$.
 
\section{The Learning-based Algorithm Design}
  To solve the computation offloading problem in Eq.~(\ref{obj}a) via DRL, we first transform the problem into a POMDP, where each mobile user $u$ is an agent interacting with the MEQC environment independently. Then, we design a multi-agent hybrid continuous-discrete DRL algorithm for learning the optimal sustainable offloading strategy without prior knowledge.
\subsection{POMDP for Quantum Computation Offloading}
\subsubsection{Observation Space} 

  We first define the observation space $O_u^t$ of mobile user $u$ at the current decision slot $t$ as a union of the local computation conditions $L_u^t \triangleq [f_u^l, s_u, n_u, Q^L_u, D^L_u]$, edge computation conditions $E_u^t \triangleq [f_u^E, Q_u^E, k_1,\ldots, k_U]$, and wireless connection conditions $W^t_u \triangleq [p_u, g_u(1), \ldots, g_u(E)]$, which is denoted as $O^t_u \triangleq \{L_u^t, E_u^t, W^t_u\}$.
  
\subsubsection{Action Space}
  Each mobile edge user $u$ as a learning agent needs to maintain a hybrid discrete-continuous action space, which is denoted as $
      A^t_u \triangleq \{a_u, \phi_u\}.
$
  Here, $a_u$ is the discrete action to determine the server that user $u$ should connect to, and $\phi_u$ is the continuous action to indicate the portion of the task that user $u$ should process locally.
\subsubsection{Reward Function}
At each environment step, each learning agent of mobile user $u$ can gain rewards $R = -C(\boldsymbol{a}, \phi)$ w.r.t. current observations and actions.
\begin{figure*}[t]
	\centering
	\begin{minipage}[t]{0.24\linewidth}
		\centering
		\includegraphics[width=1\linewidth]{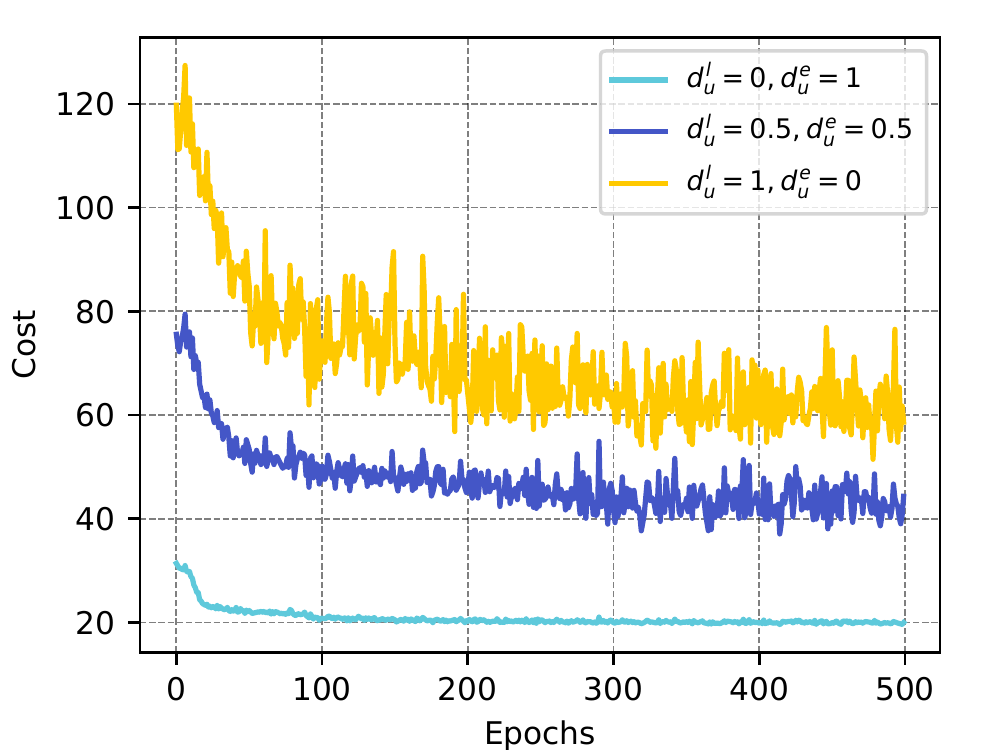}
		\caption{System cost v.s. training epochs}
		\label{fig:con10}
	\end{minipage}%
        \begin{minipage}[t]{0.24\linewidth}
		\centering
		\includegraphics[width=1\linewidth]{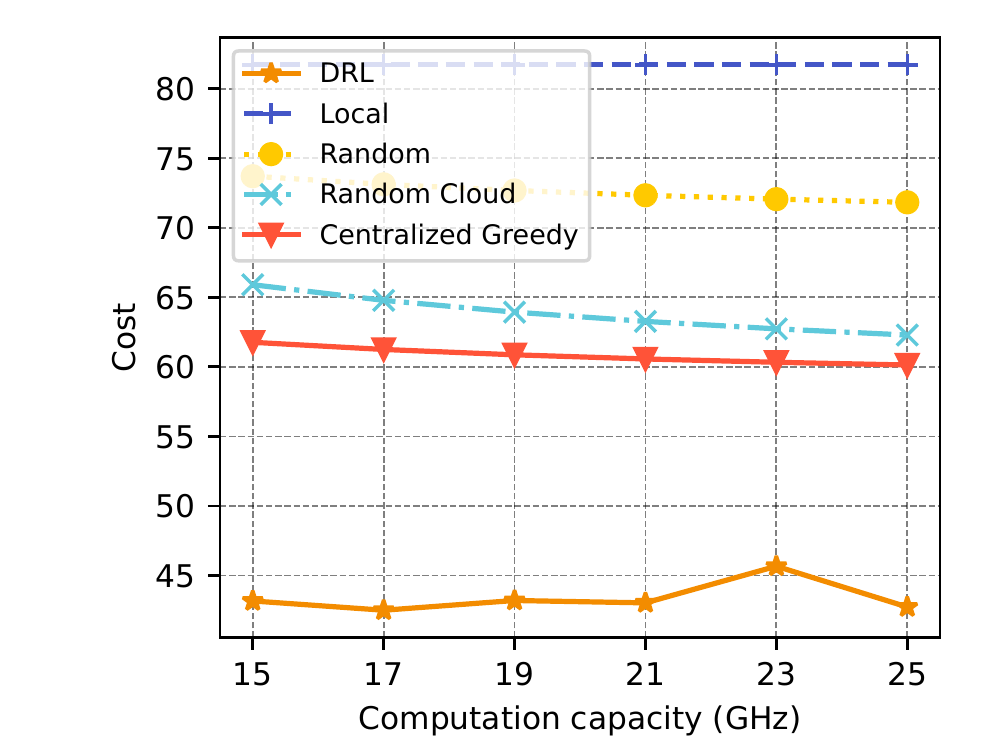}
		\caption{System cost v.s. classical computation capacity.}
		\label{fig:classical}
	\end{minipage}%
	\begin{minipage}[t]{0.24\linewidth}
		\centering
		\includegraphics[width=1\linewidth]{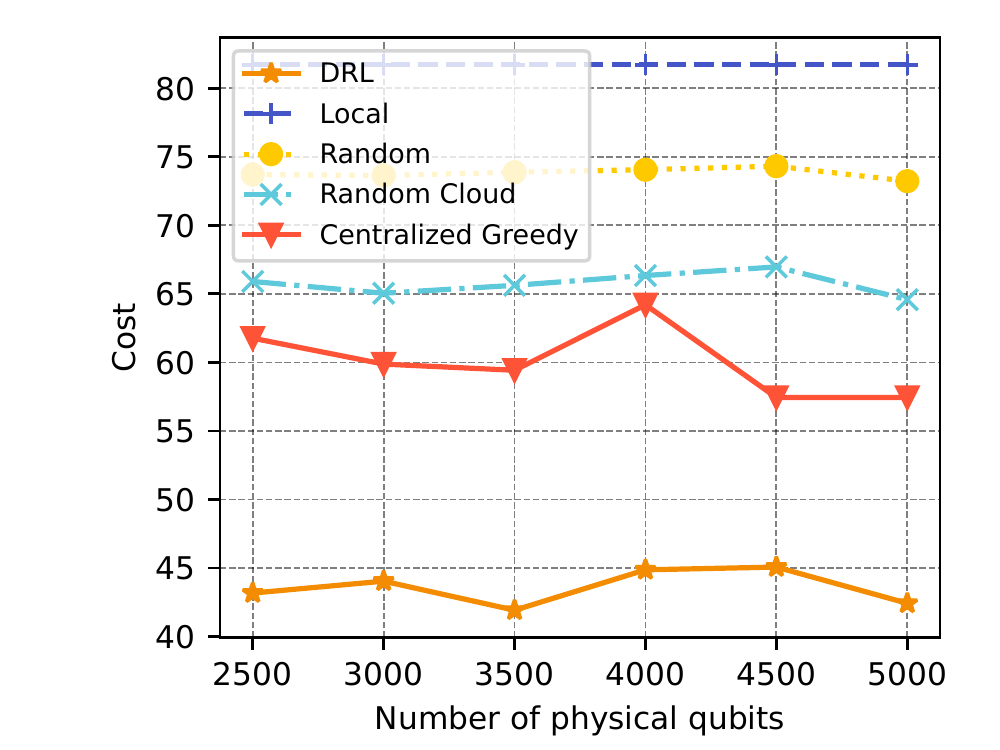}
		\caption{System cost v.s. quantum computation capacity}
		\label{fig:quantum}
	\end{minipage}
	\begin{minipage}[t]{0.24\linewidth}
		\centering
		\includegraphics[width=1\linewidth]{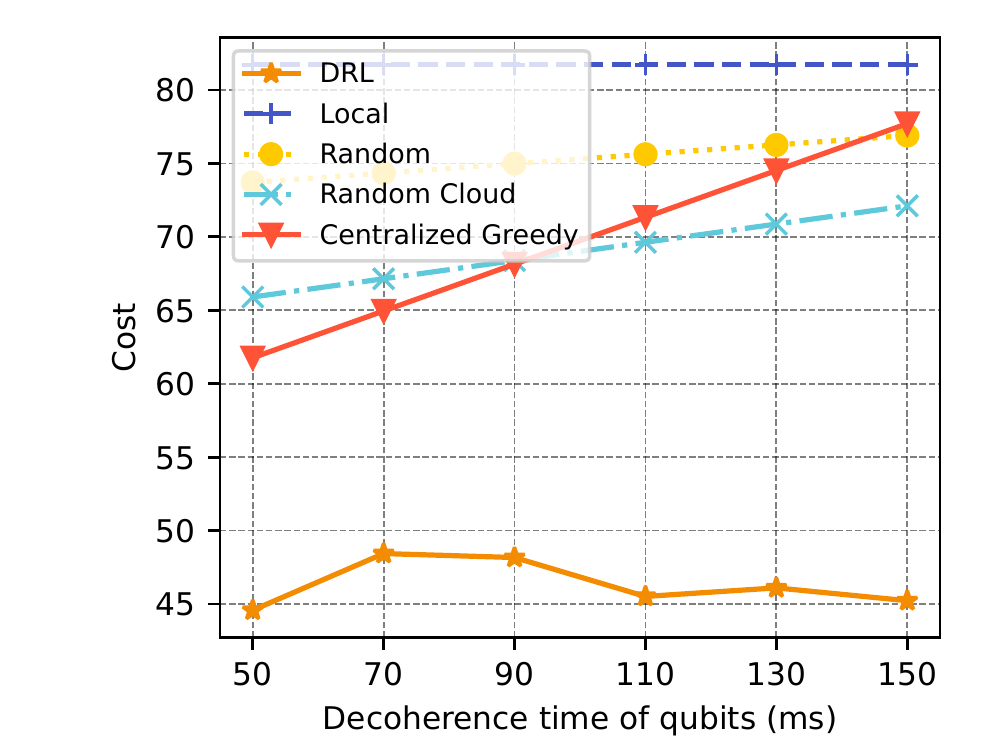}
		\caption{System cost v.s. qubit quality}
		\label{fig:qtime}
	\end{minipage}
\end{figure*}
\subsection{Multi-agent Policy Evaluation and Improvement}
Each learning agent maintains two pairs of actor-critic networks $(V^a, \pi^a)$ and $(V^\phi, \pi^\phi)$ for determining discrete and continuous actions, respectively~\cite{neunert2020continuous}. In multi-agent policy evaluation and improvement, each learning agent first collects experiences during the interaction with the environment into its replay buffer. Then, the performance of the current strategy is evaluated using the critic networks with general advantage estimation. Finally, based on the evaluation of the critic networks, the policy networks are improved via gradient ascent w.r.t. the learning rate, while the critic networks are updated via gradient descent w.r.t. the learning rate. Let $\mathbb{E}_\pi(\cdot)$ denote the expected value of a random variable given that the agent follows policy $\pi$ and $\gamma\in[0,1]$ which is the reward discount factor used to reduce the weights as the time step increases. Finally, the expected long-term value $V^a$ and $V^\phi$ are maximized, and thus the sustainability of the MEQC system is optimized.

		\section{Experimental Results}\label{sec:exp}
\subsection{Parameter Settings}

  In the simulation experiment, we first consider a MEQC system with 10 mobile users and 10 edge servers. 
  For the communication model, the channel gain of each user is randomly assigned from the set [4, 8] and the transmission power is allocated from [0.01, 0.2] mWatts. The bandwidth owned by each server is set to 20 MHz. For the classical computation model, the local computational capacity is randomly assigned from the set \{1, 2, 3\} GHz and the edge computation capacity is randomly assigned from the set \{10, 15, 20\} GHz. The chip coefficient is assigned to $\gamma=10^{-11}$ for the energy consumption per CPU cycle according to the measurement method as in \cite{chen2015efficient}. For the quantum computation model, the number of physical qubits at each edge server is randomly assigned from [1000, 5000]~\cite{resch2021benchmarking}, the error correction's concatenation level is randomly selected from [1, 2, 3], the qubit operation temperature is set to 0.1 K, the signal generation temperature is set to 300 K, and the attenuation is set to 40 dB. 
  In line with~\cite{fellous2022optimizing}, the typical qubit frequency is set to 6 GHz, the 1qb gate time is set to 25 ns, the 2qb gate time is set to 100 ns, the measurement time is set to 100 ns, the number of refrigeration stages is set to 5, the threshold for error correction is set to $2\times 10^{-4}$, the heat produced by signal generation \& readout is set to 10 $\mu$W, the heat produced at 4K by paramps is set to 10 nW, and the heat produced at 70K by HEMT amps is set to 50 $\mu$W. For each logical qubit with error correction's concatenation $k$, the number of required physical qubits is $(91)^k$, the number of physical 1qb gates is $\frac{28}{185}(64)^{k}$, the number of physical 2qb gates is $\frac{64}{185}(64)^{k}$, and the number of physical measurement gates is $\frac{28}{185}(64)^{k}$.
  
  In this paper, we focus on ray-tracing rendering tasks, where the coordinate is set to 16$\times$16$\times$16, the resolution of each frame is set to 128$\times$128, the number of frames is randomly assigned from [1024, 10240], the number of primitives is randomly assigned from \{3, 4, 5, 6, 7, 8, 9\}, the number of rays of each primitive is set to 3. Therefore, data size of each classical computational task is randomly assigned from [160, 1600] MB and the number of required CPU cycles is randomly assigned from 3 $\times$ \{$2^3,\ldots,2^9$\} cycles/byte. In addition, when the classical computational task is converted to a quantum computational task, the required numbers of qubits are $\{20,\ldots, 26\}$ and the required circuit depths are \{813,\ldots, 6560\}. For the weights of each user $u$ for both the latency and energy, they are set to 0.5 by default.

  The hyper-parameters in the proposed DRL algorithm are set as follows. The discrete and the continuous policies are parameterized by two-layer fully connected networks with 256 hidden units. We train the algorithm for 500 epochs that consists of 2000 steps in each epoch. After each epoch, the policies are updated twice with a batch size of 128, discounting factor of 0.95, and a learning rate of 0.001. We perform our experiment with Python 3.9, PyTorch 1.12.1, and CUDA 11.6.
  
  \subsection{Convergence Analysis}

  First, we analyze the convergence of the proposed multi-agent hybrid discrete-continuous DRL. In Fig. \ref{fig:con10}, we show the performance of the proposed algorithm in achieving convergence in user scenarios with different preferences. For the more energy-conscious user scenarios, the proposed algorithm takes about 50 epochs to achieve convergence performance. On the other hand, for the more delayed user scenario, the proposed algorithm takes about 300 epochs to reach the convergence performance, which is similar to that of the scenario where the delay and energy are equally weighted.
  
  \subsection{Performance Comparison}

  Then, we evaluate the proposed system model and the proposed DRL algorithm under different system settings. We use the local, random, random cloud, and centralized greedy schemes as the baseline solutions for performance comparison. In Fig. \ref{fig:classical}, for all the algorithms except the local execution algorithm, the cost of the algorithm decreases as the computation capacity of the edge server increases. However, as we can observe from Fig. \ref{fig:quantum}, the increase in the number of qubits will not significantly affect the consumption of computation offloading in MEQC. The reason is that the number of qubits at edge servers is a hard-cutting success probability to determine whether the quantum advantage can be brought to mobile users. Finally, we increase the quality of qubits (i.e., decoherence time) in quantum computers and illustrate the results in Fig.~\ref{fig:qtime}. The costs of quantum computing increase dramatically as the quality of qubits increases. Nevertheless, the proposed DRL algorithm leads to choosing the most economical offloading strategy without the impacts of increasing energy consumption from quantum computing. Overall, the proposed algorithm can reduce at least 30\% of cost compared with baseline solutions
  
		\section{Conclusions}
		In this paper, we have explored the MEQC paradigm and dynamic quantum computation offloading. In MEQC, quantum advantages are brought from edge servers equipped with quantum computation capacities to mobile users. We have formulated the dynamic computation offloading problem in MEQC as a mixed-integer programming model. To achieve decentralized decision-making for mobile users, we have formulated MEQC as a POMDP where users are agents and proposed a hybrid discrete-continuous multi-agent DRL to learn the optimal sustainable offloading strategy. Experimental results have shown that the proposed algorithm is efficient in terms of convergence and performs better than the existing baseline solutions under different system settings.
		
		
		%
		\appendix
  \subsection{The Noise Model in Quantum Computers}\label{app:noise}
  We take the common concept that stages should have equal attenuation and be regularly spaced in orders of magnitude of temperature between $T_{\tmop{gen}}$ and $T_{\tmop{qb}}$~\cite{martin2022energy}. In other words, if we want a total attenuation of $A$, we take
\begin{equation}
  A_i = A^{1 / (K - 1)}, \quad T_i = T_{\tmop{qb}} \left(
  \frac{T_{\tmop{gen}}}{T_{\tmop{qb}}} \right)^{(i - 1) / (K - 1)},
\end{equation}
for $K$ stages of cooling of quantum computers.

We consider the thermal photon contribution to the noise is reduced to an acceptable level by a chain of attenuators on the ingoing microwave line. These attenuators are kept cold by cryogenics, and hence they thermalize the signal to come down the line from hotter temperatures, reducing the population of thermal photons. For a chain of $K$ cooling stages with $K - 1$ attenuators (e.g., $K = 5$), the error probability of a physical qubit is~\cite{fellous2022optimizing}
\begin{equation}
  \epsilon_{\tmop{err}} = \frac{\gamma \tau_{\tmop{step}}}{2} \left( \frac{1}{2} + n
  (T_1) + \sum^{K - 1}_{i = 1} \frac{n (T_{i + 1}) - n (T_i)}{\tilde{A}_i} 
  \right),\label{error}
\end{equation}
where $T_1 = T_{\tmop{qb}}$, and $n(T)  = (\exp [\hbar \omega / k_B T] - 1)^{- 1}$
is the Bose-Einstein function at the qubit frequency. In Eq. \eqref{error}, we observe that the noise can always be reduced by increasing the attenuation, which results in higher power consumption. 

		\subsection{Energy Consumption of Quantum Computers}\label{app:energy}
The resource cost is defined as the power $P_{\pi}$ consumed to bring the
qubit from $|0\rangle$ to $|1\rangle$, which can be defined as $
    P_{\pi} = \frac{\hbar \omega_0 \pi^2}{4 \gamma \tau^2_{1}},$
where $\omega_0$ is the transition frequency and $\gamma^{-1}$ is the spontaneous emission rate, i.e., the decoherence time, depending on specific qubit technology.
The power consumption per physical 2qb gate averaged over the timestep of the quantum computer can be defined as
  $P_{2} = P_{\pi} \sum_{i = 1}^K \frac{T_{\tmop{gen}} - T_i}{T_i}
  (\tilde{A}_i - \tilde{A}_{i - 1})$,
where $T_{\tmop{gen}}$ is the generation temperature, $T_i$ is the intermediate
temperature at stage $i$, and $\tilde{A}_i = A_i\times \cdots \times A_2 \times A_1$ is the total
attenuation between $T_i$ and the qubits. Moreover, the power consumption per physical 1qb gate can be defined as
  $P_{1} = \frac{\tau_{1}}{\tau_{\tmop{step}}} P_{2}$,
where $\tau_{1}$ is the 1qb gate time and $\tau_{\tmop{step}}$ is the timestep of quantum computers.
Finally, the power consumption per physical qubit is
\begin{equation}
    P_{\mathrm{Q}} = \frac{T_{\text{ext}}}{T_{\text{gen}}} 
  \dot{q}_{\text{gen}} + \frac{T_{\text{ext}}}{T_{\text{hemt}}} 
  \dot{q}_{\text{hemt}} + \frac{T_{\text{ext}}}{T_{\text{para}}} 
  \dot{q}_{\text{para}},
\end{equation}
where $\dot{q}_{\text{gen}}, \dot{q}_{\text{hemt}},$ and $ \dot{q}_{\text{para}}$ are heat produced at $T_{\text{ext}} = T_{\text{gen}}, T_{\text{hemt}}=70 K,$ and $T_{\text{para}}=4 K$, respectively.
		\bibliographystyle{IEEEtran}
		\bibliography{bare_conf}

	\end{document}